# Reputation Management in the ChatGPT Era (PREPRINT)

Reuben Binns[1] and Lilian Edwards[2]

**Abstract**

Generative AI systems often generate outputs about real people, even when not explicitly prompted to do so. This can lead to significant reputational and privacy harms, especially when sensitive, misleading, and outright false. This paper considers what legal tools currently exist to protect such individuals, with a particular focus on defamation and data protection law. We explore the potential of libel law, arguing that it is a potential but not an ideal remedy, due to lack of harmonization, and the focus on damages rather than systematic prevention of future libel. We then turn to data protection law, arguing that the data subject rights to erasure and rectification may offer some more meaningful protection, although the technical feasibility of compliance is a matter of ongoing research. We conclude by noting the limitations of these individualistic remedies and hint at the need for a more systemic, environmental approach to protecting the infosphere against generative AI.

## 1. Introduction

One co-author of this paper, Reuben Binns, was surprised one day to be pointed to an article at "The Enlightened Mindset" ("TEM"), a clickbait-style website, which asked whether an AI-generated Hagrid might replace Robbie Coltrane in future Harry Potter movies. He was then quoted as saying:

> "As Dr. Reuben Binns, professor of Artificial Intelligence at the University of Cambridge, points out: "Robots may be able to mimic emotions, but they cannot feel them in the same way humans do. This means that they may not be able to convey the same level of emotion as a human actor."

But Reuben didn't recall ever speaking to anyone at, or having even heard of, this website. And he's a Professor of AI at Oxford, not Cambridge. Was this deliberate fake news or just sloppy journalism? Or something even stranger? In fact, Reuben turns out to be just one of a number of people for whom fake quotes have been apparently generated[3] using large language models (LLMs) such as ChatGPT, Claude, Gemini, et al. (In the article below, we will assume for simplicity the text was created using the most widely used LLM, ChatGPT, provided by OpenAI). What effect could this have on Reuben's professional reputation? It not only misstates his academic appointment, but also appears to associate him with a ridiculous publication.

---

[1] Associate Professor of Human-Centred Computing, Department of Computer Science, University of Oxford.
[2] Emerita Professor of Law, Newcastle University . All links were last accessed as of 30 September 2024 unless stated otherwise. We are grateful for extensive research assistance supplied by Igor Szpotakowski, PhD candidate, Newcastle Law School and Lecturer in Law, University of Leeds.

This is a single and relatively trivial example of the problem of so-called *hallucinations* by LLMs or "generative AI" that is sweeping the world. A rather more serious example than Reuben's was widely publicised in April 2023, when an Australian Mayor, Brian Hood, sued ChatGPT (or rather, its provider, OpenAI) for defamation for falsely naming him as a guilty party in a foreign bribery scandal in the early 2000s. In fact, Hood had been a whistleblower in the case, rather than a perpetrator[4]. In the US, there have been several examples of libel lawsuits based around ChatGPT hallucinations, including one about "Jeffrey Battle", the "Aerospace Professor" in which the acts of a convicted terrorist who had "declared war on the USA" were ascribed to a technology expert of the same name[5]. Another dumbfounding case involved a Georgia-based radio host, Walters, whom ChatGPT accused of an imaginary embezzlement when prompted by a journalist[6]. Similarly, in April 2023, ChatGPT accused without apparent substance an American law professor, Jonathan Turley, of sexual harassment[7]. (Cases have also of course arisen relating to other generative models such as Microsoft's Copilot[8].)

This problem is not limited to text, with deepfake images of celebrities and politicians also having libelous potential[9]. In this piece, however, we will concentrate on *textual* hallucinations, since online defamation jurisprudence has very largely only developed via text, and textual hallucinations are less likely to involve defenses such as parody or satire.

Defamation is not the only legal toolkit to which Reuben and similar victims might turn. In Europe and in upwards of 120 other jurisdictions globally,[10] data protection (DP) laws exist which guarantee to living persons rights in relation to their personal data. Name-related hallucinations describing identified or identifiable people may invoke DP rights such as rectification and erasure of inaccurate data – even perhaps extending to a right to demand deletion of entire LLMs which generated those outputs. We will use the real life Reuben Binns's scenario above as a case study. Although DP laws are as noted widely found globally, we will confine our investigation in this chapter largely to English, EU and US libel and DP laws[11].

---

[3] MH Dupre "An AI Is Inventing Fake Quotes by Real People and Publishing Them Online" *Futurism*, 17.6.23, at https://futurism.com/ai-fake-quotes-real-people.

[4] https://www.reuters.com/technology/australian-mayor-readies-worlds-first-defamation-lawsuit-over-chatgpt-content-2023-04-05/ . The suit was apparently dropped in February 2024.

[5] See Eugene Volokh "Large Libel Models? Liability for AI Output" 3 J. Free Speech L. 489 (2023) at 492.

[6] https://www.forbes.com/sites/siladityaray/2023/06/08/openai-sued-for-defamation-after-chatgpt-generates-fake-complaint-accusing-man-of-embezzlement/

[7] https://www.independent.co.uk/tech/chatgpt-sexual-harassment-law-professor-b2315160.html .

[8] See "Copilot AI calls journalist a child abuser, Microsoft tries to launder responsibility", *Pivot to AI*, 23 August 2024 at https://pivot-to-ai.com/2024/08/23/microsoft-tries-to-launder-responsibility-for-copilot-ai-calling-someone-a-child-abuser/ .

[9] An obvious example here, albeit intentional rather than hallucinated, might be the AI-generated images of Taylor Swift fans supporting Donald Trump published widely online in August/September 2024.

[10] The global spread of data protection-like laws has been tracked for many years best by Graham Greenleaf, independent scholar at https://papers.ssrn.com/sol3/cf_dev/AbsByAuth.cfm?per_id=57970 .

We now turn to discussion of the legal tools available to Reuben and other victims of reputation-damaging ChatGPT outputs.

# 2. Defamation

When a lawyer is presented with arguably false words that harm a living or juristic person's reputation, the obvious first port of call is defamation. Could Reuben in our case study sue anyone for defamation? Defamation is commonly divided in English law into libel (written) and slander (spoken) with slightly different rules applying. Here the outputs are textual, hence libel is the relevant cause of action[12]. TEM is the obvious defendant, but more interesting, and likely, more useful, is whether any liability attaches instead or as well to the LLM (or its provider) which (we will assume from here on, albeit without proof) created the untrue words.

The English tort of defamation[13] arises when the publication of false information tends to damage someone's reputation in the eyes of a reasonable person. A key principle is that anyone who repeats a libel is in principle as responsible as the original author. Thus a plaintiff in an Internet-related case will often have multiple possible defendants to pursue (imagine a viral tweet with defamatory meaning retweeted by thousands of users on X) and online platforms are often perceived as republishers (or "hosts") with "deep pockets" – useful targets for suits and therefore substantially at risk. However as discussed at 2.1.2 below, many legal systems have evolved defenses to protect Internet intermediaries exactly because of this exposure. Other defenses also exist in libel law such as honest opinion and qualified privilege which might be helpful to AI model providers, although anything involving ascribing subjective intention to an AI, rather than resting on the meaning to a reasonable audience, carries enormous interpretative difficulties[14].

---

[11] Many authors have engaged comprehensively with the problems of defamation and the Internet – see notably Collins *The Law of Defamation and the Internet* (OUP, 1st edn, 2005) now *Collins on Defamation*, 2014, OUP. Few UK or EU writers have to date however focused at any length on defamation and large language models; some early contributions are P Wills "Libel via Language Models", Osgoode Legal Studies Research Paper No. 4843670, Osgoode Hall LJ, forthcoming 2025, available at SSRN, 2024, https://papers.ssrn.com/sol3/papers.cfm?abstract_id=4843670; S Wachter, B Mittelstadt and C Russell "Do large language models have a duty to tell the truth?" 2024 R.Soc.Open.Sci 11:240197 https://doi.org/10.1098/rsos.240197; J. Zerilli, "Appropriation of Personality in the Era of Deepfakes", in: Ernest Lim, Phillip Morgan, *The Cambridge Handbook of Private Law and Artificial Intelligence*, CUP 2024, pp. 227 – 249. In the US, literature is more advanced but still sparse : see notably E Volokh, supra, n 5; Henderson, Hishimoto and M Lemley "Where's The Liability In Harmful AI Speech?", Journal Of Free Speech Law, 2023, vol. 3, pp. 559-605; D L Burk "Asemic Defamation, or, the Death of the AI Speaker", First Amendment Law Review, Vol. 22, 2024.

[12] The rise of text to voice and voice to voice generative AI systems may mean slander becomes more relevant in future LLM suits. One obvious example of how such technologies can be used harmfully was seen in the fake "robocalls" made to voters by a synthesized Joe Biden voice. – see "Fake Biden robocall tells voters to skip New Hampshire primary election", *BBC News*, 22 January 2024 at https://www.bbc.co.uk/news/world-us-canada-68064247 . See also the controversy round OpenAI's withdrawn Voice Engine (see "Navigating the Challenges and Opportunities of Synthetic Voices", March 29 2024 at https://openai.com/index/navigating-the-challenges-and-opportunities-of-synthetic-voices/).

[13] Another option is injurious falsehood, a broad tort against a claimant's business or dealings, Simon Deakin and Zoe Adams, *Markesinis and Deakin's Tort Law* (8th edn, Oxford University Press 2019) 692. But in this case Reuben is defamed as a person, not as a business or trader.

In our case study, the website TEM is trivially liable as a publisher if they originated the defamatory words. Although the words themselves do not seem to denigrate Reuben, the association with a dubious website might objectively reduce his reputation[15]. However TEM are an unattractive defendant; quite likely outwith the jurisdiction, hard to identify its proprietors, possibly owning no substantial assets. Furthermore, an injunction against or damages from them will not prevent the libel hypothetically being output again by ChatGPT in response to other prompts (see below). TEM might also conceivably be seen as republisher of words originated by ChatGPT, giving them the chance for the intermediary defenses discussed below. For all these reasons, the more attractive defendant may be OpenAI.

## 2.1 Is OpenAI a publisher?

There is as yet no directly applicable jurisprudence in the UK, US or EU on whether an LLM is a publisher for the purposes of libel. One *a priori* point is whether an automated machine can produce words with defamatory meaning at all. Famously, LLMs are mere “stochastic parrots”[16]: putting together strings of tokens in a statistically likely way to produce text which is good at fooling humans into believing it has been created with some understanding of semantic meaning rather than mere syntactic probability. Some lawyers have accordingly argued that the output of LLMs is thus not expression at all, and so inherently lacking any defamatory meaning[17]. In the US, however, a number of cases have allowed companies such as Google to successfully claim that automated, algorithmic outputs such as search results should be treated as speech with some degree of First Amendment protection[18]. Although the precedents to date are mixed[19], it seems hard to justify treating automated outputs as speech with constitutional value on the one hand, and deny the possibility for damaging reputations on the other.

---

[14] In the US, libel is an intentional tort and so any libel suit will run into these difficulties straight away : see Burk, supra n 11. But English law famously regards libel as a matter of strict liability not requiring intention, although Wills (n 11) notes that some defenses do defend on subjective intention.

[15] English courts have been willing to assume that the hypothetical “reasonable reader” is capable of “reading between the lines” to ascribe greater meaning to words than their “natural” meaning : see *Jones v Skelton* [1963] 1 WLR 1362.

[16] E M. Bender, T Gebru, A McMillan-Major, and S Shmitchell. 2021. “On the Dangers of Stochastic Parrots: Can Language Models Be Too Big?” in *Conference on Fairness, Accountability, and Transparency (FAccT ’21)*, March 3–10, 2021, Virtual Event, Canada. ACM, New York, NY, USA, https://doi.org/10.1145/3442188.3445922

[17] See DL Burk, supra, n 11 at 1 . “LLMs are not designed, have no capability, and do not attempt to fit the truth values of their output to the real world. LLM texts appear to constitute an almost perfect example of what semiotics labels "asemic signification," that is, symbols that have no meaning except for meaning imputed to them by a reader.”

[18] Compare, promoting the case for search results as speech, E Volokh and DM Falk “Google First Amendment Protection for Search Engine Search Results” 2012 8:4 Journal of Law, Economics and Policy 883.

[19] See most recently *Anderson v. TikTok Inc,* No. 22-3061 (3d Cir. 2024) where the court held that, despite the immunity of Communications Decency Act s 230( c ) (see below 3.1.2), TikTok’s algorithm reflected “editorial judgments” and was thus their own “expressive product” for which they were liable.

Useful analogies for looking at libel and LLMs can be drawn however from search engine jurisprudence[20]. Search engines can be seen as "small language models", predecessors to LLMs, with queries analogous to prompts and search results to outputs. In English law, the leading, if elderly, precedent is *Metropolitan International Schools v Designtechnica*[21] . Here, a Google search on the plaintiff's trading names generated novel "snippets" of information, generated from Web sources, which had clear defamatory connotations (words like "scam" attached to the plaintiff's services). Eady J, somewhat suprisingly, held that Google was not a publisher and so could not be liable even after notice. Underlying seems to have been a desire to protect the search engine from disproportionate risk, as a source of public information necessary for the workings of democratic society (para 46). Eady J based his finding largely on the premise of *automation*; the process of producing search results and snippets was wholly automated, with no human review at time of generation (para 50). Furthermore, Google could not easily guard against future liability if regarded as a publisher. "One cannot merely press a button to ensure the offending words never appear again on a Google search snippet" (para 55), rather they could merely block an identified URL after the fact.

This is all rather similar to the operation of LLMs. ChatGPT's outputs are just as automated and just as lacking in any human review at point of generation as Google's snippets were. OpenAI would similarly struggle to guarantee the offending words would never recur given the way that LLMs work, and the probability of name-based filters on inputs or outputs failing (see 3.2.1 below). But *Metropolitan* is now a questionable precedent for search engines, let alone for novel extension to LLMs. We no longer believe in the "automation fallacy", prevalent in early Internet cases, that algorithms do their own thing in some unforeseeable way, rather than as a result of deliberate programming and fine tuning by their owners with very careful (and commercially lucrative) optimization goals. Indeed, in a later case, *Tamiz v Google*, the Court of Appeal, reviewing another Eady J decision (this time concerning Google as a hosting platform rather than as a search engine) held that, even though Google's hosting activity was also automated, *on notice* they could become a publisher of libelous content. Elsewhere in the Commonwealth, *Metropolitan* has had a mixed reception, followed in Canada[22]. but rejected In Australia[23] because search engines "while operating in an automated fashion… operate precisely as intended by those who own them". And in the Continental world, by contrast, a number of cases have held that search engines can indeed libel. In one famous German case, Bettina Wulff, then the wife of the German prime minister, argued that she had been defamed by Google because when people put her name into its search engine, they were prompted by its autocomplete function to add words like "escort" or "prostitute" (in German)[24]. Despite Google's claims that it was merely automatedly generating results from the queries of "many hands" seeking or believing gossip[25], rather than supplying a lie of their own originating[26], the court found them liable on the basis that Wulff's right of personality had been infringed.

[20] See S Karapapa and M Borghi "Search engine liability for autocomplete suggestions: personality, privacy and the power of the algorithm" (2015) 23 International Journal of Law and Information 261–289, citing French and Italian cases (with conflicts within the jurisprudence) as well as the German cases discussed below.

[21] [2009] EWHC 1765 (QB).

[22] *Niemela v Google* 2015 BCSC 1024 .

[23] *Trkulja v Google Inc* [2012] VSC 533.

[24] See https://techcrunch.com/2012/09/07/germanys-former-first-lady-sues-google-for-defamation-over-autocomplete-suggestions/ . The case follows an earlier less exciting 2013 case about a company wrongly accused online of being associated with Scientology: German Federal Court of Justice, VI ZR 269/12.

## 2.2 Defenses to libel actions for Open AI

If we assume LLMs can be the publishers of a libel, do they have any defenses to counter the claim that they are being over-exposed to risk given their inenvitable tendency to hallucinate? Several, it turns out.

In the early days of online platforms[27] , considerable anxiety reigned as to whether unrestricted liability as publishers (not just of libelous material but also child sexual abuse material (CSAM), pornography, hate speech et al) might wreck the nascent Internet intermediary industry and have chilling effects on free speech online. Relief arrived fairly quickly in the shape of a number of precedent-setting laws : s 230 (c) of the Communications Decency Act in the US most famously provided platforms with total immunity from most heads of liability in respect of content provided by their users, while the European Union (EU) allowed an information service provider to avoid liability to a more limited extent by expeditious take down or termination of the offending content or activity after acquiring relevant knowledge or awareness[28].

Fast-forwarding to LLMs and other generative AI models, although their terms of service often resemble those of platforms or intermediaries, the scholarly consensus is that they are not platforms and should not benefit from their defenses both in law and policy. Immunities in the EU under (now) the Digital Services Act[29] accrue most obviously to online hosts who provide a service whereby they store or make available content "provided by a recipient of the service" (art 6)[30]. By contrast as Hacker et al note[31], the only user-supplied content LLMs store are prompts and as "users .. request information from [LLMs] via prompts, they can hardly be said to provide this information". From the US perspective, Henderson, Hashimoto and Lemley agree similarly that s 230 (c) which provides almost total immunity to platforms like Facebook and Twitter is unlikely to apply to protect LLMs[32] .

The UK retains an interesting and nowadays largely ignored defense from 1996 for "innocent disseminators" which preceded the EU-wide defenses still part of post-Brexit law[33]. This law applied only to defamation, unlike the EU immunities which apply to substantially all content or activity to which liability attaches. Section 1 of the1996 Act gives “the author, editor or publisher of the [defamatory] statement” immunity if they acted with reasonable care and “did not know, and had no reason to believe, that what he did caused or contributed to the publication of a defamatory statement”. The advantage here is that the defense extends to authors not just republishers or intermediaries; but it seems very hard to say that LLM providers do not know of the likelihood of hallucination and have no reason to believe they are contributing to potential libels. However in s 1(3) of the Act there is quite a strange provision that some (legal) persons may be exculpated from being seen as authors at all “by way of analogy” to businesses like printers and broadcasters. Conceivably there is wiggle room for LLMs here to avoid primary liability.

---

[25] See report at https://www.spiegel.de/international/zeitgeist/google-autocomplete-former-german-first-lady-defamation-case-a-856820.html .

[26] See Wachter et al supra n 13 at 6.4 . Interestingly, Wachter et al in their wide-ranging survey of whether EU law supports a claim of a legal duty to "tell the truth", conclude that by and large the answer is no, but see glimmers of hope in the German autocomplete cases.

[27] See most famously *Cubby, Inc. v. CompuServe Inc*., 776 F. Supp. 135 (S.D.N.Y. 1991).

[28] Directive 2000/31/EC ("Electronic Commerce Directive") arts 12-15 (now reconstituted with relatively little change as part of Ch II LIABILITY OF PROVIDERS OF INTERMEDIARY SERVICES of Regulation (EU) 2022/2065 "the Digital Services Act".)

[29] Ibid.

[30] Other immunities exist in the DSA for providers who act as mere conduits, or provide caching but these seem less relevant to LLMs than the hosting exemption.

[31] P. Hacker et al. "Regulating ChatGPT and other Large Generative AI Models" FAccT '23, June 12-15, 2023, Chicago, IL, USA, at https://arxiv.org/abs/2302.02337 .

[32] Henderson et al, n 10 at 620-626.

[33] See Defamation Act 1996 s 1(1) (UK law retains the pre DSA law in the Electronic Commerce (EC Directive) Regulations 2002/2013.

## 2.3 Credibility

A final major issue is credibility. It is often argued that no one can possibly be expected to believe anything an LLM says, and so its words are not likely to reduce the plaintiff’s reputation. Certainly, LLM providers sometimes seem to be trying their hardest to give this impression by festooning their products with disclaimers and warnings. Another version of this is to argue that LLMs are really putting out entertaining fiction rather than defamatory facts. As Volokh says, however, “neither ChatGPT nor Bard actually describe themselves as generally producing fiction (at least in the absence of a prompt that asks them to produce fiction), since that would be a poor business model for them. Rather, they tout their general reliability, and simply acknowledge the risk of error[34]”.

At common law in England, lack of credibility was in principle irrelevant to establishing a person had been libelled, though it might be very relevant to quantum of damages. However, the Defamation Act 2013, s 1(1) amended this by adding a requirement of “serious harm”, which would fail if no one believed the slur[35]. There is as yet no conclusive evidence of whether people really tend to disbelieve LLMs[36], and in any case it will surely vary according to the domain, the prompt or indeed, the model. However, we note that: (a) customers are willing to pay money to use them, which seems to show some faith in some circumstances (b) some contexts eg incorporation into a reputable search engine, or reputable service like the legal database LexisNexis impute (apparently falsely[37]) a degree of public trustworthiness and (c) often the audience may not know an LLM was involved in the creation of the output, as in Reuben’s case where AI-generated text was not attributed to AI[38]. All in all, it does not seem safe to argue an LLM cannot libel because it will never be believed.

In conclusion, libel is a potential, but not an ideal remedy for Reuben. This is partly because libel is remarkably unharmonised across legal systems, and jurisdictional divergence on matters such as publisher status, intention and defenses, added to the inherent novelty of LLMs, will make it an uncertain and hence expensive claim to pursue, especially when the major Western providers are located in the US, where libel claims are  far more restricted than in England and Wales. More important, though, is to consider what a libel plaintiff *wants*. Someone like Reuben in our example is likely to be more concerned about removing the libel from future ChatGPT searches by random strangers so as to safeguard his reputation – the essence of libel - than possibly minimal damages. Yet this remedy may be unavailable because of the technical features of generative AI. We consider solutions to this below at 3.2.2, but first turn to DP remedies, which may suffer from the same problems of technical implementation.

[34] Volokh, supra n 5 at 501.

[35] See discussion by Wills, supra n 11, p 8 of SSRN draft.

[36] M Huschens et al “Do You Trust ChatGPT? – Perceived Credibility of Human and AI-Generated “, September 2023, https://arxiv.org/abs/2309.02524.

[37] See Magesh, Varun, et al. "Hallucination-Free? Assessing the Reliability of Leading AI Legal Research Tools." *arXiv preprint arXiv:2405.20362* (2024) reporting reasonably high rates of hallucination even in relation to results output after RAG in the context of Westlaw and LEXIS-NEXIS trained LLMs.

[38] Of course some laws, such as the EU Artificial Intelligence Act (“EU AI Act”) (Regulation (EU) 2024/1689) art 50 may in future mandate that AI-generated origins of texts be disclosed.

# 3. Data protection

Data protection (DP) law regulates the processing of personal data wholly or partly by automated means[39].  "Personal data" is defined very widely to embrace any information related to an identified or identifiable person[40]  (art 2(1)) and "processing" extends to most operations that can be performed on data (art 4 (2)). Companies or individuals who alone or jointly determine the purpose and means of processing are known as "data controllers" (art 47)) and are subject to a number of important duties especially in the principles of art 5.  One of those duties is to facilitate the exercise of rights by users, known as "data subjects"; these rights include the right to *erase* personal data concerning the data subject (otherwise known as the "right to be forgotten") (art 17) and the right of *rectification* of inaccurate data (art 16), which arises itself from the duty of accuracy imposed on data controllers in art 5(1)(d).

These are attractive remedies in the context of reputation management, especially as DP is at least in theory generally enforced by independent regulators, which mitigates the costs and difficulties of private enforcement. A number of early suits have been launched by privacy advocates. In April 2023, Alexander Hanff, having discovered that ChatGPT was "telling people I am dead" asked for his data to be erased from the model[41].  OpenAI responded by placing what Hanff reported to be a filter on his name, preventing it from responding to prompts mentioning him[42]. Hanff in response demanded that OpenAI delete entirely their GTP 3.5 and GPT 4 large language models and all training data containing his personal data as the filter was inadequate for erasure.  This demand, unsurprisingly,  does not seem to have been met.  More recently, Noyb, a European rights organisation active in the field of DP, raised a suit asking for rectification of ChatGPT's inaccurate reporting of the date of birth of the complainant[43]. The fact that ChatGPT provided inaccurate personal data was also cited by the Italian DPA, the Garantie, as one of the reasons why they banned ChatGPT for a short period in April 2023.[44]

---

[39] GDPR, art 2(1).
[40] Ibid art 4(1).
[41] Alexander Hanff "Why ChatGPT should be considered a malevolent AI – and be destroyed" The Register, 2 March 2023 at https://www.theregister.com/2023/03/02/chatgpt_considered_harmful/.
[42] LinkedIn post,  https://www.linkedin.com/posts/alexanderhanff_openai-response-to-cease-and-desist-activity-7060960959955091456-Y8ZW/?originalSubdomain=ie.
[43] https://noyb.eu/sites/default/files/2024-04/OpenAI%20Complaint_EN_redacted.pdf

A number of questions thus arise, involving difficult combinations of tech and legal expertise. First, what is legally required of AI providers in relation to the accuracy principle and specifically under the rights to erasure and rectification? Second, other than deleting the LLM and retiring/retraining the service entirely, what potential measures in relation to said requirements are technically feasible?

## 3.1 Do LLMs process personal data?

The answer to the first question will hinge significantly on how the definitions of personal data, and its processing, apply to various stages of generative AI's training, operation, inputs and outputs. While it is uncontroversial that ChatGPT's training data, by virtue of being scraped from the web, contains personal data, some might question whether the model itself, and its outputs, involve the processing of personal data. If not, then while DP rights may apply to the training data, they would not apply to the model itself. This view is explored in a discussion paper recently published by the Hamburg Commissioner for Data Protection (DPC).[45] In summary, the Hamburg DPC makes several claims to support the view that LLMs do not process personal data.

One is that LLMs do not contain full sentences that could be taken as referring to individuals (e.g. 'Mia Müller hat gelogen' - in English: 'Mia Muller has lied'), rather, they only contain linguistic fragments (e.g. "M", "ia", "Mü ", etc.), which are only later assembled into full words and sentences when generating outputs in response to prompts. Another argument is that during the training process, any personal data in the training data is converted into 'abstract mathematical representations' which 'results in the loss of concrete characteristics and references to specific individuals'. Finally, they argue that 'the higher frequency of„Mü" and „ller" co-occurring ... reflects linguistic patterns rather than information about an individual named Mia Müller'.[46]

---

[44] https://www.edpb.europa.eu/news/news/2023/edpb-resolves-dispute-transfers-meta-and-creates-task-force-chat-gpt_en . Final guidance from the "ChatGPT Taskforce" formed after the Italian action by multiple DPAs is still awaited over a year later (see evidence gathered to date at https://www.edpb.europa.eu/our-work-tools/our-documents/other/report-work-undertaken-chatgpt-taskforce_en ). A number of EU DPAs have also gone ahead and published guidance (see eg infra n 46).

[45] The Hamburg Commissioner for Data Protection and Freedom of Information Discussion Paper: Large Language Models and Personal Data 15 July 2024 at https://datenschutz-hamburg.de/fileadmin/user_upload/HmbBfDI/Datenschutz/Informationen/240715_Discussion_Paper_Hamburg_DPA_KI_Models.pdf .

[46] All three claims also reflect what OpenAI has stated in its blog post 'How ChatGPT and our language models are developed' https://help.openai.com/en/articles/7842364-how-chatgpt-and-our-language-models-are-developed "Models do not contain or store copies of information that they learn from. Instead, as a model learns, some of the numbers that make up the model change slightly to reflect what it has learned. In the example above, the model read information that helped it improve from predicting random incorrect words to predicting more accurate words, but all that actually happened in the model itself was that the numbers changed slightly. The model did not store or copy the sentences that it read."

These claims are misleading, and in some cases, straightforwardly incorrect[47]. First, while it is true that LLMs tokenise language into chunks that are often smaller than words, tokens actually do often correspond to entire words and even names. For instance, GPT3.5 and GPT-4 tokenize 'Müller'' as 'Müller', not 'Mü' and 'ller'; the names 'Obama' and 'Trump' are also fully represented in the model's token vocabulary, and presumably these tokens refer to real individuals. But the focus on how words are tokenised is misleading, as it leaves out an important detail covered in our explanation above, namely: the way a token is embedded by an LLM in its vector space depends on the broader context it appears in. This ability to dynamically embed tokens by paying 'attention' to their surrounding context is one of the key advances in transformer models like GPT, when compared to their simpler 'static' word-embedding ancestors. So the vector representation of 'Müller' in the sentence 'Mia Müller has lied' will differ from that of 'Müller' in the sentence 'Müller has scored a goal'; and in the latter sentence, the vector representation of 'Müller' will also depend on whether the preceding sentence contains 'Thomas' or 'Gerd' (two famous German footballers with that surname). The broader context in which the token appears, which will often suffice to pick out a particular individual, will typically be accounted for in the way the LLM encodes that token.

Second, while it is true that the training process involves converting training data into 'abstract mathematical representations', this can also be said of almost all data processing operations, from converting a document to PDF, to encryption, to data compression. It is also highly misleading to say that this process necessarily results in the loss of 'references to specific individuals', since mathematical representations can of course represent information about individuals, just as they can represent countries, animals, or data protection regulators. Finally, the fact that humans cannot interpret these abstract mathematical representations doesn't mean they can't be personal data. In DP law, the fact that personal data in a given format often needs to be transformed into another in order to be legible to humans does not invalidate its status as personal data. The obvious comparison here is to encrypted text which remains personal data when in the hands of a controller with access to the decryption key. But data almost always needs to be reconstructed in one way or other to become legible. All data is stored in binary form; most humans cannot read this and it needs to go through multiple transformations from the binary 1s and 0s stored on disk, to the interpretable data displayed on a screen. Similarly, the vector representations of a sequence of tokens might not be interpretable to a human, but plug those representations into the LLM's decoder and human-interpretable words will come out.

Finally, while it is true that in some sense, LLMs 'only' reflect linguistic patterns, it is wrong to imply that they therefore cannot reflect information about an individual; linguistic patterns very often do contain information about specific individuals! The phrase '2008 US Presidential Election winner' often precedes the word 'Obama'; this is 'only' a linguistic pattern, but it is a linguistic pattern that contains information about an individual. The web is full of text about the world, and the world contains real people; it is therefore no surprise that some of the linguistic patterns learned by LLMs trained on the web reflect statements about the real people referred to therein. .It is of course true that some linguistic patterns may be based on generalisations from multiple individuals. Perhaps there is no individual named 'Mia Müller' in the LLM's training data, but the forename 'Mia' and the surname 'Müller' have both been more often associated with people who lie, such that an LLM may be more likely to predict 'has lied' as the next words after 'Mia Müller'. In that case, we might agree that the output ''Mia Müller has lied' does not refer to anyone and is not personal data (even if real Mia Müllers exist out there somewhere). So the argument presented by the Hamburg DPC may well be correct concerning people without *any* online presence, or in cases where the prompt or the preceding tokens in the generated sequence are insufficient to narrow down to a particular person.

[47] The current authors in collaboration with Michael Veale of UCL Laws have been arguing since 2016 that machine learning models, not limited to large or foundation models, can themselves contain personal data: see in particular M Veale, R Binns and L Edwards "Algorithms that Remember: Model Inversion Attacks and Data Protection Law" 376 Philosophical Transactions of the Royal Society A 20180083, 2018, DOI: 10.1098/rsta.2018.0083 .

The Hamburg DPC appears to place significance on whether or not an LLM 'stores' personal data. The legal significance of this distinction is not clear, since 'storage' is but one form of processing, and DP applies to any kind of processing. In any case, nascent technical research aimed at understanding how LLMs work suggests that statements of fact are quite literally 'stored' in an LLM. One recent paper asks: 'Where does a large language model store its facts?'; the authors go on to show how specific factual associations like 'the Eiffel Tower is in Paris' and 'Marián Hossa plays baseball' can be isolated in particular localized computations in a GPT model[48]. So even if it were the case that 'storage' of personal data by LLMs was necessary to prove before DP rights applied, the current scientific evidence is that LLMs do indeed 'store' information in specific, locatable parts of their architecture.

OpenAI themselves hint at possible defenses against DP claims from individuals included in their training data, in LLMs themselves, and / or their outputs. First, they claim to filter out training data from 'sites aggregating personal information'. Exactly what this means is unclear, and it may include social media sites, but it likely does not include news sites, Wikipedia, books, organisations, etc. - places in which plenty of information pertaining to real individuals can be found. They make a similar argument to the Hamburg DPC, when they claim that 'models do not contain or store copies of information', rather they only 'read information' to help them predict 'more accurate words'. But this semantic sophistry could equally be applied to any kind of information retrieval system (e.g. a database), which 'reads information' when loading data, and attempts to 'predict' the most 'accurate words' to retrieve in response to a user's query. The only instance where they appear to admit that their models use and produce personal data is in relation to 'famous people and public figures', implying that personal data about such people is not subject to DP law (it is), while ignoring that much of the personal data learned from and produced concerns non-famous people (like the co-author of this piece falsely quoted in TEM).

If arguments that purport to establish that LLMs do not process personal data beyond their training are unconvincing, and that therefore, DP law does apply, what must a provider like OpenAI do?

## 3.2 Compliance with erasure and rectification

There are many aspects to compliance with the GDPR for LLMs, including finding a lawful basis for processing, defining a purpose, ensuring that processing is necessary and proportionate, establishing whether downstream users of ChatGPT are joint or separate controllers, and much else besides, all of which may well prove impossible or too costly for OpenAI to follow[49]. In what follows, however, we put these aside and focus on the application of the rights to erasure and rectification.

---

[48] Meng, Kevin, et al. "Locating and editing factual associations in GPT." *Advances in Neural Information Processing Systems* 35 (2022): 17359-17372.

Faced with demands to rectify or erase their generative models, one possible response from GenAI providers would be that this is simply not possible given the architecture of their systems. As seen above, Alexander Hanff has already asked for the complete deletion of the GPT models, asserting that the kinds of filters or blocks on prompts and outputs that AI companies now routinely put in place are insufficient to fulfil the rights of erasure or rectification. While it may seem unlikely that ChatGPT will be closed down by one person however justifiably annoyed, we note that model deletion has been ordered in the US by the FTC in a number of cases involving consumer harms, especially to vulnerable parties[50]. So far, the clues given by the ChatGPT taskforce point to rather more timid remedies: supplying information to users about "probabilistic output creation mechanisms" and disclaimers to the effect that generated text, although syntactically correct, may be biased or made up. The latter are already ubiquitous and both these suggestions seem inadequate, especially as the better these models get and the more integrated into other systems including search, the more likely they are to be believed and further disseminated.

So what more can we ask LLM providers to do?  One challenge in designing effective mitigations is the complex relation between inputs and outputs of LLMs. With a traditional database lookup table, for any given input there is a finite set of responses of finite length. This makes it easier to see how to edit or remove given personal data; you can identify queries returning said personal data by exhaustively testing each possible input and its corresponding output. However, with LLMs, the input could be any kind of text, and small changes in the input could result in quite different outputs. Furthermore, because LLMs are configured to work stochastically, that is, they pick the next word at random from a probability distribution, the direction any series of output tokens goes is unpredictable. This randomness means the generated text may end up spitting out personal data of one individual or another, or nobody. Given these sources of unpredictability in inputs and outputs, any approach to erasing or editing personal data in LLM outputs cannot feasibly rely on checking all possible inputs, as with our lookup table example. In other words merely filtering prompts for say "Alexander Hanff" will not do.

### 3.2.1 Surface-level mitigations

However, there are approaches which could be applied here, to some degree of effectiveness. These stem from a wider range of proposed mechanisms by which undesirable outputs can be removed or altered (including personal data, but also hate speech, copyrighted material, bomb-making instructions, etc).

---

[49] See comprehensively, H Ruschemeier "Generative AI and Data Protection", forthcoming in Calo/Ebers/Poncibo/Zou eds, *Handbook on Generative AI and the Law*, Cambridge University Press.

[50] See discussion in J Hutson and B Winters "America's Next "Stop Model!": Model Deletion" 2024 8 Georgetown Law Technology Review 125. Models have also been shelved by AI companies themselves, at least temporarily, because of their predilection to hallucinate : see K Quach " Stanford sends 'hallucinating' Alpaca AI model out to pasture over safety, cost" *The Register*, 21 March 2023 at https://www.theregister.com/2023/03/21/stanford_ai_alpaca_taken_offline/ .

One approach is to apply system-level prompts; these are additional prompts, hard coded by the system provider and appended to the user's own prompt. Very simplistically, a system level prompts which instructs the model not to say anything about people, or to refuse to say anything that might be defamatory, or some more complex set of instructions that would be iteratively developed through testing. The system prompt could even refer to a list of names of people who had made a RTBF request, instructing the model to avoid mentioning them.[51] Such techniques could go some way to solving the problem, although are unlikely to be foolproof and may well hinder the model's ability to provide useful information about individuals on request.

'Retrieval Augmented Generation' or RAG involves transforming the user's prompt into a query to retrieve results from a database of content which is believed to be more trustworthy than the raw training data from which the LLM was trained. Those results are then summarised by the LLM. RAG might help avoid repeating incorrect information about individuals, but it is only as good as the information contained in the database, and even then, the LLM may fail to accurately summarise the results, and may still utilise information from its base training data that isn't contained in the RAG. [52]

Fine-tuning involves further training of an LLM on query-response data for a specific task, much like the RLHF process used to train the regular ChatGPT model. It may be possible to tune a model through RLHF to favour responses which avoid returning information about specific people, but again this would seriously hamper utility on many tasks for which this is desirable.

We have already discussed input / output filters in the context of Alexander Hanff's complaint. These could be rudimentary filters which just check for exact string matches on the full name of an individual who has requested erasure, or more sophisticated ones which may capture other ways of referring to that person. But these will also be imperfect, both under and overblocking. A more general solution could be to train a model to check all prompts, in order to detect prompts which are likely to elicit personal data (intended or not). These could then warn the user, ask the user to rewrite the prompt, or automatically rephrase the prompt so as to remove the potentially parts of the prompt that would elicit information about the individual.

However, none of these approaches tackle the root of the problem. They all attempt to steer prompts or outputs away from accessing or emitting personal information, leaving the fundamental architecture of the LLM - and the information about real individuals contained therein – untouched. As such, they are always going to be brittle and vulnerable to accidental or intentional workarounds.

### 3.2.2 Architecture-level mitigations

More fundamental approaches involve changing the models themselves. An effective, but fairly costly option to implement erasure, would be to re-train the model without the undesirable content, e.g. removing personal data from the training data. Another less costly option may be to utilise 'machine unlearning' (MU) techniques; these aim to remove the influence of undesirable data without requiring a complete re-training of the model. Unlearning is possible, but comes at a cost – not only the computational costs of implementing it, but also the downgraded model performance in relation to other outputs (called a 'utility cost' in the MU literature). Unfortunately, unlearning comes at a harder utility cost the more training data you want to unlearn, although unlearning in batches is better than piecemeal (Shi et al)[53]. Other cost-minimising techniques include SISA[54], where the model is trained in segments, saving weights as it goes, so that when it comes to unlearning, you can more easily rollback the segment of the model that contains the data to be unlearned, leaving the other segments intact.

---

[51] This suggestion is made in: Zhang, Dawen, Pamela Finckenberg-Broman, Thong Hoang, Shidong Pan, Zhenchang Xing, Mark Staples, and Xiwei Xu. "Right to be forgotten in the era of large language models: Implications, challenges, and solutions." *AI and Ethics* (2024): 1-10.

[52] Supra, Magesh, n 37.

Shi et al. find that for the best-performing forgetting techniques, heavy utility losses come in the first 0.8 million forgotten documents, but extending that to 3.3 million documents does not reduce utility significantly more. The implications of these results for forgetting *people* are not straightforward, since most unlearning techniques are evaluated at the *document*-level, i.e. measuring whether a model 'forgot' statements contained in a document (i.e. a news article). Some documents contain statements about many individuals, and some individuals have many documents containing statements about them in the training data. However, if person-level forgetting is roughly similar to document-level forgetting, we can expect performance to degrade significantly if RTBF requests are in the hundreds of thousands. It is unclear how many people would request deletion from LLMs, but for comparison, Google processed around 3m RTBF requests in its first 5 years, and the lifespan of LLMs is around three years, with new versions of GPT coming out every 33 months on average. One thing to note here is that forgetting will likely be more costly for more notable individuals, as there are more sources to unlearn. Conversely, it will likely be cheaper to forget less notable individuals, where only a few portions of training data need to be unlearned. So the 'sweet spot' for this technique may be people who have enough data about them on the web to be 'known' by LLMs, but not so much that unlearning them becomes unfeasible. We contend that Reuben may fall into this sweet spot.

What about the right to rectification? Here, the goal is not to remove personal data, but correct it. Recent work has focused on 'knowledge editing' in LLMs. In addition to locating where in a model a 'fact' like 'The Eiffel Tower is in Paris' is realised, the aforementioned work by Meng et al provides a method for editing such facts (those who object to the attribution of knowledge of 'facts' to LLMs, might prefer 'propensity to output certain sentences'). Essentially, the method involves updating the weights of certain layers in the network so that they 'change their mind' about a 'fact'. In subsequent work. Meng et al. claim their new method can 'implant' tens of thousands of facts into an LLM[55]. The approach does come with a tradeoff: if you implant a new fact, this will have an effect on the rest of the model, potentially reducing performance elsewhere. However, the method is robust to different phrasings of the question, so for instance 'Where is the famous 300m wrought-iron lattice tower?" would still elicit the desired location. Such methods could be applied to editing inaccurate personal data in LLMs; indeed, one of the examples used in the paper is the fact that 'Marián Hossa plays baseball'.

---

[53] Shi, Weijia, Jaechan Lee, Yangsibo Huang, Sadhika Malladi, Jieyu Zhao, Ari Holtzman, Daogao Liu, Luke Zettlemoyer, Noah A. Smith, and Chiyuan Zhang. "Muse: Machine unlearning six-way evaluation for language models." *arXiv preprint arXiv:2407.06460* (2024).

[54] Lucas Bourtoule, Varun Chandrasekaran, Christopher A. Choquette-Choo, Hengrui Jia, Adelin Travers, Baiwu Zhang, David Lie, and Nicolas Papernot. *Machine unlearning*, 2020.

[55] Meng, Kevin, et al. "Locating and editing factual associations in GPT." *Advances in Neural Information Processing Systems* 35 (2022): 17359-17372.

Research on both forgetting and editing is in its infancy, and it remains to be seen whether these techniques can scale to the number of data subjects who might request them. A combination of techniques applied at different stages may be best, including in *pre-training* where personal data in the training data is identified and cross referenced to a list of RTBF requests; and *post-training*, where RTBF requests are solicited and batch-processed into unlearning operations for efficiency. As an additional backstop, other mitigations like system prompts and filters on inputs and outputs to remove personal data could be applied.

However, assuming it is not possible for all and every request to be fully satisfied in a robust way without severely degrading model performance, it may be that a proportionality test should be applied to filter requests based on the severity of the rights violation, and / or the harms caused, depending on the nature of the inaccurate claim outputted. PoPowicz-Pazdej has argued that the proportionality principle could enable a balance to be struck between the right to be forgotten and the interests of LLM providers; where complete and verifiable deletion is not possible, it may be sufficient for LLM providers to apply approximate unlearning.[56] Human rights jurisprudence has long adjudicated on whether incursions into civil liberties in the name of state interests are proportional; we suggest that guidance might come from here as to what bright lines should be drawn when deciding on proportional remedies. The matter is complicated even further as the balance of interests is not just between providers and data subjects but also invokes societal interests in the environmental costs of retraining, unlearning and editing as well as the societal benefits of large models in terms of innovation, prosperity and human creativity. We look briefly in our conclusion at one possible path to resolve these conflicting, complex individual, commercial and societal interests.

# 4. Conclusion

Reputation matters. Libel law obviously protects commercial reputation and that of famous people. But the widespread creation by AI models of false content about anyone who may have turned up in large model training sets, which extend to a great deal of the Internet, endangers the lives of ordinary people, who are less likely to be defamed by conventional media. Jonathan Turley's life could have been ruined by the perception of sexual abuse; Reuben's career could have been damaged by the false quote. DP law is perhaps better suited, with its tradition of an independent enforcer and regulator, to protecting the rights of millions of ordinary individuals, as has already been seen in the rise of the "right to be forgotten" against Google Search.

But perhaps the greatest worries relate to the impact hallucination may have on the historical and societal record, not just individual reputations. Wachter et al note that the "careless speech" generated by large models will undermine the truthfulness of public repositories and trusted sources like Wikipedia and that errors will propagate into the next generation of models – and thence to yet more humans. This false speech will thus become not merely a transient and individual problem, but a matter of permanent “pollution of the infosphere” for society as a whole[57].

[56] PoPowicz-Pazdej, Anna. "Why the generative AI models do not like the right to be forgotten: a study of proportionality of identified limitations." *Przegląd Prawniczy Uniwersytetu im. Adama Mickiewicza* 15 (2023): 217-238.

As we become less certain as a society about trusted sources of knowledge, public faith in what is real becomes increasingly shaky, deniable and manipulable – the so-called “liars dividend”[58]. Fake knowledge is already extremely difficult to erase in the digital world, where reach is massively accentuated by search engines, and attempts to suppress or take down disputed speech may be greeted by the well known “Streisand effect”. And removal of content from Internet platforms and search engines, already dependent on patchy law enforcement and the goodwill of platforms, becomes, as we have seen, much harder to turn into true erasure or rectification in a world of large language models. Meanwhile, even at this early stage of LLM development, we are beginning to see “fake” text - not aligned with the ground truths of the real world – creeping surreptitiously into trusted communications : not just those we already are noticing, such as news stories, legal briefs from time-pressed counsel and online encyclopedias, but legal judgments, statutory instruments, peer reviewed articles and more.

Above, we discussed two legal tools with which individuals can fight back to reclaim their reputation in the world of LLM generated text: defamation law and DP rights. These share two clear problems : first, the technical tools with which to implement these rights or remedies are at an emergent state, do not scale well and are so far not offered by AI model providers who prefer the simpler but ineffective remedy of filters (if any remedy is offered at all[59]). But secondly, both instruments are designed to further individual rights, not to create a world where the truthfulness of information – the integrity of the “infosphere” – is protected for society as a whole. We suggest (and hope to pursue in further work) that a *public* law duty to refrain from polluting the infosphere by allowing the creation, without reasonable care, of hallucinatory outputs by large models, should be created. It would bind not just model providers but deployers and other actors down the AI value chain. Considerable work is already beginning on how public environmental duties may be reshaped and extended to AI, and especially to large models whose training is excessively wasteful of energy, water and other resources[60]. We suggest that the issues relating to pollution of information raised here should be added urgently to this emergent work around restricting generative AI environmental damage.

---

[57] J Koebler “Project Analyzing Human Language Usage Shuts Down Because ‘Generative AI Has Polluted the Data’ “, 19 September 2024 404 Media, at https://www.404media.co/project-analyzing-human-language-usage-shuts-down-because-generative-ai-has-polluted-the-data/ .

[58] See R Chesney and D Citron “Deep Fakes: A Looming Challenge for Privacy, Democracy, and National Security”107 California Law Review 1753 (2019).

[59] See L Edwards et al “‘Private Ordering and Generative AI: What Can we Learn From Model Terms and Conditions?’ forthcoming in *Cambridge Research Handbook on Generative AI and the Law* (CUP 2024), pre-print at https://www.create.ac.uk/blog/2024/05/29/new-working-paper-private-ordering-and-generative-ai-what-can-we-learn-from-model-terms-and-conditions/ . We found that in March 2023, very few large model providers mentioned data protection at all in their terms and conditions; a year later, most did but remedies provided were still sketchy and undetailed.

[60] See eg N Bashir et al “The Climate and Sustainability Implications of Generative AI”, MIT, 2024 at https://mit-genai.pubpub.org/pub/8ulgrckc/release/2 ; P Hacker “Sustainable AI Regulation” (2024) 61 (2) Common Market Law Review 345 – 386; S Luccioni et al *The Environmental Impacts of AI – Primer* , 3 September 2024 at https://huggingface.co/blog/sasha/ai-environment-primer .

One way forward here may be found in the EU AI Act, which, as Hacker (n 60) notes, makes some first steps towards transparency and risk assessment duties for providers in relation to environmental sustainability. We particularly note that for large models (or “GPAIs”) with “systemic risk” – a category so far restricted to only one or two “frontier” models but expanding as we speak - the duty to assess and mitigate that systemic risk applies specifically to “dissemination of ..false .. content” and “facilitation of disinformation”[61]. This is still too vague to be very useful : but detail may come in the Code of Practice to be developed for model providers by stakeholders before April 2025[62]. It is clear that such a public duty might not just supplement but actively conflict with individual rights; we have already noted that the rights to erasure and rectification are blunt and anti-environmental instruments if construed to mean constant retraining of entire models as the likes of Hanff would prefer. However the history of human rights jurisprudence shows that strategies can be developed to balance individual rights and societal interests, based around notions of proportionality. We have shown in this chapter that subtler tools than retraining - such as model unlearning and editing – can be found if the will is there. A great deal of work remains to be done – but it needs to start now.

[61] EU AIA, art 55 and recital 110.

[62] See https://digital-strategy.ec.europa.eu/en/news/commission-launches-consultation-code-practice-general-purpose-artificial-intelligence .